\documentclass[a4paper,11pt]{article}
\pdfoutput=1
\usepackage{jheppub}
\usepackage[T1]{fontenc}

\def\reff{r_{eff}}
\def\geff{g_{eff}}

\def\ep{\varepsilon}
\def\half{\frac{1}{2}}

\def\lr#1{\left(#1\right)}
\def\slr#1{\left[#1\right]}

\def\trl#1{\textrm{Tr}\lr{#1}}

\def\CPn{\mathbb C P^n}

\def\RS2{\mathbb R\times S^2_F}

\def \be  {\begin{equation}}
\def \ee  {\end{equation}}
\def \bex  {\begin{equation*}}
\def \eex  {\end{equation*}}
\def \bea {\begin{eqnarray}}
\def \eea {\end{eqnarray}}
\def \bal {\begin{align}}
\def \eal {\end{align}}
\def\quater{\frac{1}{4}}

\def\no{\nonumber\\}

\def\bibitemm#1{\bibitem{#1}}

\def \PRD {{Phys. Rev. D\ }}
\def \JHEP {{JHEP\ }}
\def \RMP {{Rev. Mod. Phys.\ }}

\allowdisplaybreaks

\title{\boldmath Matrix model approximations of fuzzy scalar field theories
and their phase diagrams}

\author[a]{Juraj Tekel}
\affiliation[a]{Department of Theoretical Physics, Faculty of Mathematics, Physics and Informatics,\\ Comenius University, Mlynska Dolina, Bratislava, 842 48 Slovakia
}
\emailAdd{tekel@fmph.uniba.sk}

\abstract{
We present an analysis of two different approximations to the scalar field theory on the fuzzy sphere, a nonperturbative and a perturbative one, which are both multitrace matrix models. We show that the former reproduces a phase diagram with correct features in a qualitative agreement with the previous numerical studies and that the latter gives a phase diagram with features not expected in the phase diagram of the field theory.
}

\begin{document} 
\maketitle
\flushbottom

\section{Introduction}
The idea of the noncommutativity of space time has been around for quite some time \cite{snyder}. It can be viewed as a fundamental property due to a discontinuous short distance structure of the space time \cite{doplicher}, as an effective description of the theory in some regime of the parameters \cite{witten,cpn} or as a useful regularization method preserving the symmetries of the continuous space time \cite{num14panero1}.

Regardless of the motivation, understanding the properties of the noncommutative field theories is essential. And the properties are strikingly different from the properties of their commutative counterparts \cite{noncom2,string}. At the core of this difference is the nonlocality of the noncommutative theories and the difference stays even when the noncommutativity is removed, due to the phenomenon of UV/IR mixing \cite{uvir2}.

In this paper, we will try to improve the understanding of one of such properties. Namely the phase structure of the scalar field theory on the fuzzy sphere . It has been show that it is different from the structure of the commutative theory and this difference has been observed in numerous numerical simulations of the phase diagram \cite{panero}. Even though there has been considerable effort to explain and derive the properties of this phase diagram \cite{ocon,samann,ydrinew1}, for example the location of the triple point, no satisfactory treatment is available yet.  We will make a couple of steps in this direction, with a nonperturbative treatment of all three phases of the theory in the whole parameter space.

We start with the description of the most important properties of the scalar field theory on the fuzzy sphere. We review its phase structure and describe two different matrix models that approximate its properties. We then proceed with the analysis of these two models. After we review the basic matrix models tools, we describe and check our approach on some known results for one of the models. The main part of our work is then the analysis of the asymmetric regime of this model and its comparison with the symmetric regime. We will find the phase transition lines and the triple point and we will compare it with the numerical results. We are more brief with the description of the treatment of the second model, as most of what we do is a straightforward generalization of the previous case. This model will turn out not to describe the phase structure of the field theory accurately.

Most of the equations we encounter will be transcendental and we will not be able to solve them in any analytic fashion. We thus retreat to numerical treatment in of these equations. We want to stress that what is treated numerically are the equations describing the models and not the models or the phase diagrams themselves.

\section{Fuzzy field theory}\label{sec2}
In this section we introduce the basic concepts of the scalar field theory on the fuzzy sphere. We also show how this theory can be described by a random matrix model and we give two different matrix models which approximate the theory in two different ways.

\subsection{The fuzzy field theory ... }

In this paper, we shall discuss only the case of the standard scalar field theory on the fuzzy sphere \cite{fuzzy2}. We start with the relations
\be \hat x_i \hat x_i = \rho^2 \ , \ \hat x_i \hat x_j - \hat x_j \hat x_i=i \theta \ep_{ijk} \hat x_k\ ,\label{sf2cord}\ee
which define the coordinates on the fuzzy sphere. These can be realized by
\be \hat x_i=\frac{2r}{\sqrt{N^2-1}} L_i \ , \ \theta=\frac{2r}{\sqrt{N^2-1}} \ , \ \rho^2=\frac{4r^2}{N^2-1} j(j+1)=r^2\ ,\ee
where $L_i$'s are the three generators of the $SU(2)$ in the $N$ dimensional representation. Since the scalar field is an arbitrary polynomial in $\hat x_i$'s, it is a general hermitian ${N\times N}$ matrix. The fuzzy sphere is a finite mode approximation to the usual sphere.

The analogues of the derivatives are then the commutators with $L_i$, the integrals become traces and with the analogy with the commutative theory, we can define the field theory by an action
\begin{align}
S(M)&=\frac{4\pi R^2}{N}\trl{-\half [L_i,M][L_i,M]+\half rM^2+V(M)}=\no&=
	\trl{\half M[L_i,[L_i,M]]+\half rM^2+g M^4}\label{eq23}\ ,
\end{align}
where we have written the case of the quartic potential relevant for our study. We have also absorbed the overall volume factor into the definition of the matrix and the coupling constants. The observables are defined by the correlation functions
\be\label{2correlators}
\left\langle F\right\rangle=\frac{\int dM\, F(M)e^{-S(M)}}{\int dM\,e^{-S(M)}}\ .
\ee
The important property of such field theory will be its phase structure. In the commutative case, the theory has two phases, disorder phase and nonuniform order phase, characterized respectively by a vanishing and nonvanishing expectation value of the field  \cite{jaffe}. It has been shown, that the noncommutative theory has an extra phase of uniform order \cite{sauber}. The field does not oscillate around any given value, breaks the translational invariance and its appearance is the consequence of the nonlocality of the noncommutative theory.

Moreover, this phase survives the commutative limit and is still present in the phase diagram. This is due to the UV/IR mixing, since the commutative limit of the noncommutative theory is very different from the commutative theory itself.

This phenomenon has been observed in numerous numerical studies, where the phase diagram of the theory and its properties have been studied for the fuzzy sphere \cite{num04,num07,num09,num14} and other noncommutative spaces \cite{fuzzydiscnum,rsfnum,num14panero2}. The diagram is characterized by the triple point, where the boundaries of the three phases meet. The numerical simulations have located this point in the interval of the quartic coupling
\be
g_c\in(0.125,0.15)\ .
\ee
As mentioned in the introduction, reconstructing this value in an analytical fashion is our main goal. Several other works have been after the same goal \cite{ocon,samann,ydrinew1}. We will comment on these results in section \ref{sec32}, where the crucial difference from the presented approach will be clear.

\subsection{.. as a random matrix model}

The field in the fuzzy scalar field theory becomes a hermitian ${N\times N}$ matrix and the quantities of interest are given by (\ref{2correlators}), which are matrix expectations values. This means that the field theory is described by a random matrix model, with the probability measure given by the action of the theory \cite{steinacker2}. We will say more about random matrix models in the section \ref{sec31}. Here, we will investigate what kind of matrix model the field theory becomes.

If we diagonalize the matrix $M\to U \Lambda U^\dagger$, the mass and potential terms in (\ref{eq23}) will not dependent on $U$ and only the kinetic term will. So we define\footnote{The factor of $N^2$ will become clear shortly.}
\be\label{31anfintegral}
I=\int dU\,e^{-N^2\half\trl{M[L_i,[L_i,M]]}}=e^{-N^2 S_{eff}(M)}\ ,
\ee
which is a function of eigenvalues only. Once we know this kinetic term effective action, we can treat the corresponding model
\be S(M)=S_{eff}(M)+\half r \trl{M^2}+g\trl{M^4}\ee
by the standard matrix model techniques. Note, that the limit of a large matrix size $N$ corresponds to the commutative limit of the field theory, as then $\theta\to0$ (\ref{sf2cord}).

The first treatment of the kinetic term effective action is due to \cite{steinacker2}. In \cite{poly}, it has been shown that it can be written in the following way
\be\label{32sefffull}
S_{eff}=\half F(t_2)+a_1 t_3^2+(b_1+b_2 t_2)(t_4-2t_2^2)^2+c_1(t_6-5 t_2^3)(t_4-2 t_2^2)\ldots=\half F(t_2)+\mathcal R\ ,
\ee
where
\be t_n=\frac{1}{N}\trl{M-\frac{1}{N}\trl{M}}^n\ ,\ F(t)=\log\lr{\frac{t}{1-e^{-t}}}\ ,\ee
and where the remainder $\mathcal R$ vanishes for moments of the semicircle distribution. This result is based on a result for the free model ${g=0}$, which can be solved exactly. As an approximation, we will neglect the remainder term and write
\be\label{2polymodel}
S(M)=\half F(t_2)+\half r \trl{M^2}+g \trl{M^4}\ .
\ee
The symmetric version of this model $\trl{M}=0$ has been analyzed the original work \cite{poly}, the full model was perturbatively analyzed in \cite{mojnovsi}. One of the goals of this work is to extend this analysis to the nonperturbative realm.


Other works have approached the integral (\ref{31anfintegral}) perturbatively, expanding the integrand and using different methods to calculate the resulting integrals \cite{ocon,samann} and most recently \cite{saman2}. In this work it has been shown that the kinetic term effective action has up to fourth order in the kinetic term a form consistent with (\ref{32sefffull}) and that the remainder $\mathcal R$ in (\ref{32sefffull}) starts with
\be\label{sec2saman}
\mathcal R=-\frac{1}{432}t_3^2-\frac{1}{3456}\lr{t_4-2t_2^2}^2\ .
\ee
The matrix model describing the field theory can be written as
\be
S(M)=\half r\, \trl{M^2}+g\, \trl{M^4}+\frac{1}{2}F(t_2)-\frac{1}{432}t_3^2-\frac{1}{3456}\lr{t_4-2t_2^2}^2\ ,\label{prematuresaman}
\ee
where one could take only the first four terms of the expansion of $F(t)$ to be consistent in the perturbative degree.

Let us note, that for both the actions (\ref{2polymodel}) and (\ref{prematuresaman}) to have the required scaling $N^2$, it has been necessary to scale the matrix and the parameters of the theory. So the parameters $r$ and $g$ in these expressions differ from the ones in (\ref{eq23}) by some factor of $N$. It is reassuring that the required scaling is the same across all the cited works and agrees with the scaling obtained by the numerical simulations.

The two actions (\ref{2polymodel}) and (\ref{prematuresaman}) are going to be the approximations we will study in this work. We will refer to them as the second moment multitrace model and the fourth moment multitrace moment, for the obvious reasons The difference between them is that the former takes into account only the first and the second moment of the matrix $M$, but they are packed in an orderly way which is not perturbative. The latter approximation includes also the third and the fourth moment but in a wild perturbative fashion. And as we will see, this difference is going to have rather dramatic consequences.

Before we proceed, let us mention that even though we will discuss only the case of the standard scalar field theory on the fuzzy sphere, some parts of what we do could be generalized to a more general theory on a more complicated fuzzy space. By introducing a different kinetic term in (\ref{31anfintegral}), we could modify the theory. The procedure to compute the second moment effective action $F$ is then straightforward, and it has been done in \cite{mojnovsi} for the case of the fuzzy $\CPn$. Also theories with no UV/IR mixing can be described as modifications with a more complicated kinetic term so it should be in principle possible to analyze them by these tools. However, it is not clear how to derive the fourth moment effective action for the modified theories.

\section{The second moment multitrace model - the symmetric regime}\label{sec3}
This section deals with the symmetric case, where we put ${\trl{M}=0}$. Most of the results in this case are know, but we will use this opportunity to illustrate our method on a rather simple model. But before we do that, we will start with a review of the large $N$ limit of the hermitian matrix models and the emergence of the phase structure. We will first deal with the simpler case of single trance models and later consider the multitrace models.

\subsection{Matrix models and the phase diagram of the single trace matrix model}\label{sec31}

Let us have a general hermitian matrix model \cite{saclay} given by the probability measure ${P(M)=e^{-S[M]}}$ and the normalized expectation values of matrix functions $f$
\be \left\langle f\right\rangle=\frac{1}{Z}\int dM e^{-S(M)}f(M)\ ,\label{matav}\ee
with $Z=\int dM e^{-S(M)}$.

In the large $N$ limit, for all the terms to contribute, we require that the probability measure $P(M)$ is of the same order as the integration measure $dM$, i.e. $e^{N^2}$. We thus require the action to scale as $N^2$ and we will write $P(M)=e^{-N^2 S(M)}$.\footnote{As we did in (\ref{31anfintegral}).}

If the action $S(M)$ as well as the function $f$ in (\ref{matav}) are invariant under a $U(N)$ similarity transform $M\to U M U^\dagger$, we can diagonalize the matrix $M$ and turn the integration over $dM$ into an integration over the $N$ eigenvalues $x_i$ of the matrix $M$ and an integration over the angular part $dU$. The Jacobian of such transformation is the Vandermonde determinant
\be
dM=dU\Bigg(\prod_i dx_i\Bigg)\Bigg(\prod_{i<j}(x_i-x_j)^2\Bigg)\ .
\ee
Since nothing in (\ref{matav}) depends on $U$, the integration over $dU$ is trivial and gives a numerical factor which is not going to be important in what follows.\footnote{Clearly, this is the point where the fuzzy field theory is different and one is left with a nontrivial angular integral (\ref{31anfintegral}).} Moreover, the Vandermonde determinant can be exponetiated into the action to obtain a problem written purely in the terms of the eigenvalues. The expectation value (\ref{matav}) becomes
\be \left\langle f\right\rangle=\frac{1}{\tilde Z}\int \lr{\prod_{i=1}^N dx_i} e^{-N^2\slr{S(x_i)-\frac{2}{N^2}\sum_{i< j}\log|x_i-x_j|}}f(x_i)\ .\label{mataveigen} \ee
The Vandermonde determinant thus corresponds to an effective repulsive potential between the eigenvalues.

The next step in the analysis of the large $N$ limit of the model is the following crucial observation. The $N^2$ scaling of the action means, that the contributions of the eigenvalue configurations with large $\slr{\ldots}$ in (\ref{mataveigen}) are going to be suppressed. Eventually, only the configuration $\tilde x_i$, which minimizes this expression\footnote{The expression is usually called the effective action in the literature. However we have already called a different quantity an effective action, so we will refrain from using this term in this context.} survives and contributes to (\ref{mataveigen}). This means that in the large $N$ limit
\be \left\langle f\right\rangle=f(\tilde x_i)\ \label{matavlargeN} \ee
and that the configuration $\tilde x_i$ of the eigenvalues is given by the saddle point condition
\be \frac{\delta S}{\delta \tilde x_i}=r \tilde x_i+4g\tilde x_i^3=\frac{2}{N}\sum_{i\neq j}\frac{1}{\tilde x_i-\tilde x_j}\ ,\label{sec3_saddle}\ee
where we have written out the explicit form for the quartic model \cite{brezin}
\be\label{3effsingle}
S(M)=\half r \trl{M^2}+g \trl{M^4}\ .
\ee
This equation is solved by introducing the eigenvalue density $\rho(x)$ and the resolvent $\omega(x)$
\be \rho(x)=\frac{1}{N}\sum_{i=1}^N\delta(x-\tilde x_i)\ ,\ \omega(x)=\frac{1}{N}\sum_{i=1}^N\frac{1}{x-\tilde x_i}\ .\ee
We expect these two functions to become continuous in the large-$N$ limit. The moments of the distribution are then given by
\be\label{momentss}
c_n=\int dx\, x^n\rho(x)=\frac{1}{N}\trl{M^n}\ .
\ee
Moreover the resolvent has a branch cut in the complex plane along the support of $\rho(x)$ and the value of $\rho(z)$ is given by the discontinuity of $\omega(z)$ by
\be
\rho(z)=-\frac{1}{2\pi i}\slr{\omega_0(z+i\ep)-\omega_0(z-i\ep)}\ .
\ee
To find $\omega(z)$, one has to make an assumption about the shape of the support of the distribution. Under the one cut assumption, where the eigenvalue distribution is supported over a single symmetric interval ${(-\sqrt \delta,\sqrt\delta)}$ the solution for the eigenvalue distribution is
\begin{subequations}\label{sec3_expressions1}
\be
\rho(x)=\frac{1}{2\pi}\lr{r+2\delta g + 4 g x^2 }\sqrt{\delta-x^2}\ \label{sec3_expressions1a},
\ee
with $\delta$ and the second moment given by
\begin{align}
4=&3 \delta^2 g+\delta r\ ,\label{sec3_expressions1b}\\
c_2=&\frac{1}{4}\delta^3 g+ \frac{1}{16}\delta^2 r\ .\label{sec3_expressions1c}
\end{align}
\end{subequations}
Under a two cut assumption, where the distribution is supported on two symmetric intervals ${(-\sqrt{D+ \delta},-\sqrt{D-\delta})}$ and ${(\sqrt{D- \delta},-\sqrt{D+ \delta})}$ we obtain
\begin{subequations}\label{sec3_expressions2}
\be
\rho(x)=\frac{2g|x|}{\pi}\sqrt{\big(\delta^2-(D-x)^2\big)\big(\delta^2-(D+x)^2\big)}\ ,
\ee
with $\delta,D$ and the second moment given by
\begin{align}
4 D g+r=&0\ ,\\ \delta^2 g=&1\ ,\\
c_2=&D\delta^2 g\ .\label{sec3_expressions2c}
\end{align}
\end{subequations}
We see, that for $r<-4\sqrt g$ the solution (\ref{sec3_expressions1a}) becomes negative and does not have an interpretation of an eigenvalue distribution. For $r>-4 \sqrt g$ the condition ${D-\delta>0}$ fails for the two cut solution. At this value, the two solutions coincide. So we find that at the line
\be\label{trafobrezin}
r(g)=-4\sqrt g
\ee
the model enjoys a phase transition from a one cut solution to a two cut solution. To see that this truly is a phase transition, we define the free energy
\be\label{3freenergy}
\mathcal F=-\frac{1}{N^2}\log Z=S(\tilde x)=\int dx\,\rho(x)V(x)-2\int dx dy\rho(x)\rho(y)\log|x-y|\ .
\ee
This quantity has a discontinuous second derivate at the line (\ref{trafobrezin}) and thus we obtain a phase transition of the third order \cite{brezin}. The free energy is also going to be useful if we have more than one stable solution possible at given values of the parameters $r,g$. Then, due to the fluctuations of the eigenvalues at a finite $N$, we can assume that the solution with the lower free energy, i.e. the more probable one, will be dominant in the large $N$ limit.

Let us note that there is a third possibility of an asymmetric one cut solution \cite{shimishimi}, which exists if $r<-2\sqrt{15 g}$. This solution however has always a higher free energy and is not realized in the large $N$ limit.

We will drop the tilde over the eigenvalues denoting the equilibrium configuration in the rest of the paper, hopefully not causing any confusion.

\subsection{Phase diagram of the multitrace matrix model and the scalar field on the fuzzy sphere}\label{sec32}

The models with the action containing also powers of the traces of the matrix $M$ are refereed to as multitrace models. For example the action for the fuzzy field theory matrix model (\ref{2polymodel}) or (\ref{prematuresaman}) is quite clearly a multitrace one, due to (\ref{momentss}). This translates into a more complicated saddle point equation (\ref{sec3_saddle}), as the variation of multitrace terms will leave explicit moments of the distribution in the equation. For a general discussion of multitrace matrix models, see \cite{mojeakty}. Here, we will not be too general and we will consider the symmetric model (\ref{2polymodel}).  

Varying this action with respect to $x_i$ yields the saddle point equation\footnote{The general variation is $\delta c_n^m/\delta x_i=mc_n^{m-1}n x_i^{n-1}$.}
\be F'(c_2)x_i+r x_i+4 g x_i^3=2\sum_{i\neq j}\frac{1}{x_i-x_j}\ .\label{3_themodel}\ee
We will treat the second moment in this equation as a parameter and require that the eigenvalue distribution we obtain yields a correct second moment. This translates into a selfconsistency condition on $c_2$. 

The saddle point condition (\ref{3_themodel}) can be written as
\be \reff x_i+4 g x_i^3 =2 \sum_{i\neq j}\frac{1}{\tilde x_i-\tilde x_j}\ ,\ee
where we have defined the effective mass parameter
\be \reff=r+F'\lr{c_2}\ .\label{3reff}\ee
The model is thus equivalent, in the large $N$ limit to a single trace matrix model (\ref{3effsingle}) with an effective mass, which depends on the second moment of the distribution. Such model has then either the one cut solution (\ref{sec3_expressions1}) or the two cut solution (\ref{sec3_expressions2}), with $r\to \reff$. Which of these solutions is realized depends on the value of $g$, namely whether $-4\sqrt{g}$ is greater or smaller than $\reff$, which is yet to be determined.\footnote{The asymmetric one cut solution is not relevant for our purposes, as it is never the preferred solution in the symmetric regime.} Conditions (\ref{sec3_expressions1c},\ref{sec3_expressions2c}) then become more complicated equations for $c_2$, the advertised selfconsistency condition.

At the phase transition (\ref{trafobrezin}) we have $\reff=-4\sqrt g$ and this simplifies the conditions considerably. We obtain $c_2=1/\sqrt g$ and then
\be\label{22trafo}
r(g)=-4\sqrt{g}-F'(1/\sqrt{g})=-5\sqrt{g}-\frac{1}{1-e^{1/\sqrt{g}}}\ .
\ee
This result was first obtained in \cite{poly}. It also shows why perturbative treatment of the previous works fails to describe the phase diagram near the origin correctly. This formula has an essential singularity at $g=0$. Since the perturbative treatment of $F(c_2)$ is an expansion in the powers of $c_2$, which in turn is equal to $1/\sqrt g$ at the phase transition, any perturbative approximation to $F$ will fail to accurately describe the phase transition line (\ref{22trafo}) close to the origin. And any strange behavior of the perturbative phase transition line close to origin is not a sign of a triple point, but sign of the nonanalyticity of the transition line. And since this is a generic feature of the matrix models describing the scalar field on the fuzzy sphere, it casts a shadow of a doubt on the second model we want to study.

We would like to study the model away from the transition line. However, the selfconsistency conditions are however impossible to solve here since the equations become transcendental. We will thus employ a different strategy to study the phase structure of the model (\ref{3_themodel}). We will work our way backwards, taking the effective parameter $\reff$ as the starting point. The distribution for the value $r$ of the original multitrace model is then given by the expressions (\ref{sec3_expressions1},\ref{sec3_expressions2}) for the value of $\reff$ such that
\be r=\reff-F'(c_2)\ .\label{sec3_refff}\ee
We will scan through the phase diagram of the effective single trace model (\ref{3effsingle}) by choosing values of $\reff$ and $g$. Since the second moment of the distribution in the single trace model is known, we can solve (\ref{sec3_refff}) numerically and find the corresponding value of $r$.  This way, we always land in some "random" point in the phase diagram of the original theory.

Points of the original phase diagram are mapped to the points of the phase diagram of the multitrace model and we will refer to this as a deformation of the original phase diagram to the multitrace one. It is very important to note, that for a general function $F(c_2)$ this transformation needs not to be one-to-one, bijective or injective and later we will come across some rather drastic changes to the diagram.

\begin{figure} [tbp]
\centering 
\includegraphics[width=.8\textwidth]{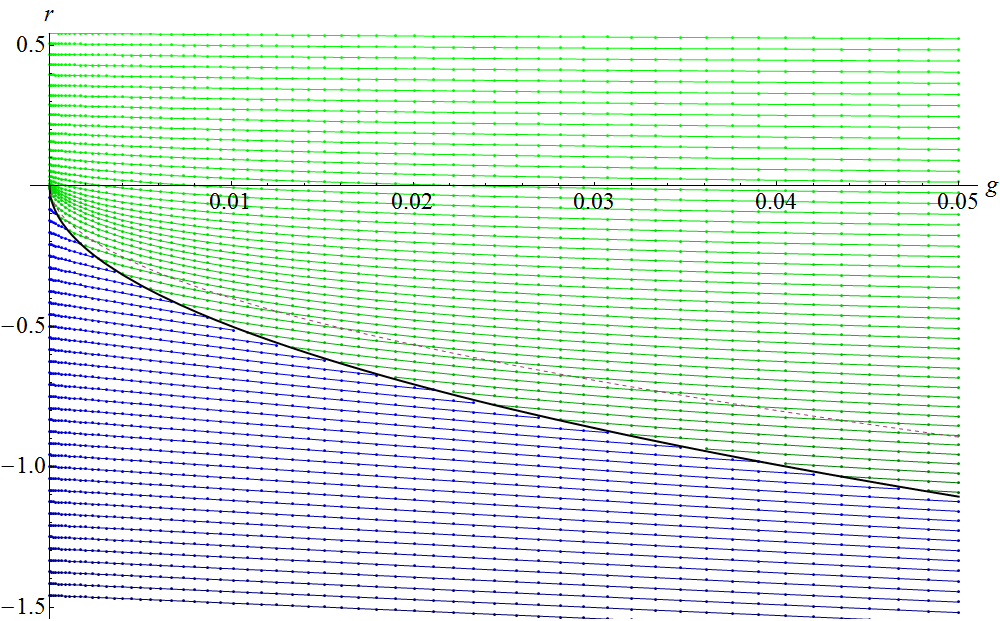}
\caption{The phase diagram of the symmetric multitrace matrix model (\ref{2polymodel}). The green region denotes the one cut solution, the blue region the two cut solution. The lines of constant $\reff$ are connected and the black line is the analytical phase transition (\ref{22trafo}). The dashed line is the phase transition of the unreformed model.}\label{fig1}
\end{figure}

The figure \ref{fig1} shows the phase diagram of the multitrace model given by the effective action (\ref{2polymodel}). The lines of a constant $\reff$ are shown to emphasize the deformation of the diagram. The green color represents the one cut region and the blue color the two cut region. We also show the original and the deformed phase transition lines as given by the analytical formulas (\ref{trafobrezin},\ref{22trafo}) respectively. We can see that the boundary between the blue and green regions nicely coincides with the analytical formula.

Since at every point we know also the eigenvalue distribution of the model, we can compute the free energy straightforwardly from (\ref{3freenergy}). This is a nonperturbative result and is the main advantage of our approach, together with being able to see the deformation of the phase diagram from the single trace to the multitrace model. Let us recall that at the phase transition, the formulas simplify and equations can be solved exactly. However once we move away from the phase transition line, we will not be able to solve the conditions (\ref{sec3_expressions1},\ref{sec3_expressions2}) anymore and these numerical tools are the only nonperturbative way to analyze the theory.

\subsection{Some other models}\label{sec3_3}

Before we proceed further, let us illustrate the numerical method on one more example, since a similar model has been studied before. Namely the model with $F(c_2)=h c_2^2$ and a fixed $r=-1$ in \cite{shishanin}. The action is thus
\be S(M)=-\half \trl{M^2}+g \trl{M^4}+ h \lr{\trl{M^2}}^2\ \ee
and the phase structure in the $(g,h)$ plane was considered. It is very easy to analyze this structure using the above numerical approach.

\begin{figure} [tbp]
\centering 
\includegraphics[width=.8\textwidth]{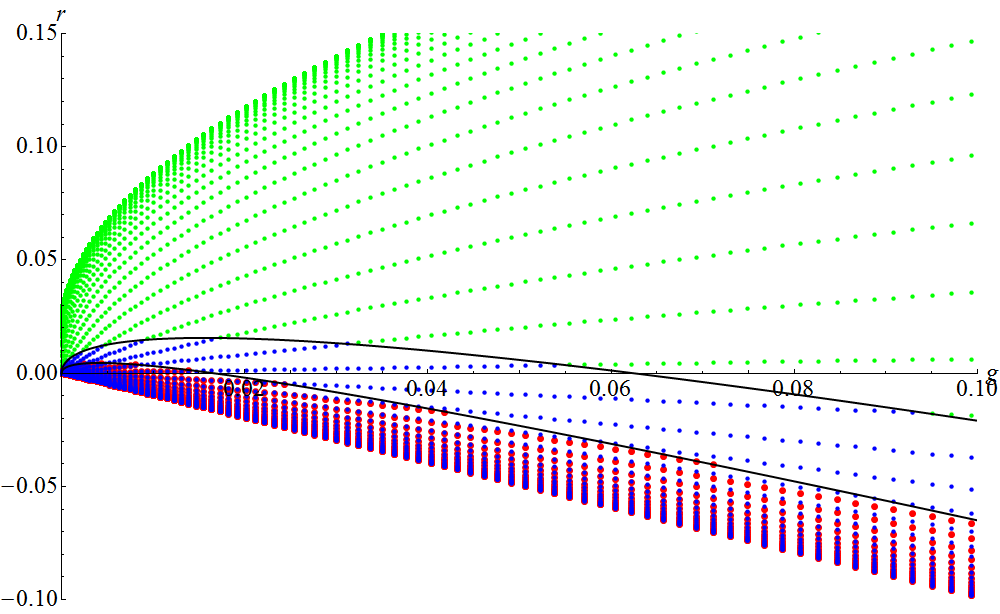}
\caption{The phase diagrams of model $S=-\half \trl{M^2}+h \trl{M^2}^2+g \trl{M^4}$. The green dots denote the existence of the one cut solution, the blue dots of the two cut solution and the red of the asymmetric one cut solution. The two transition lines are the analytical expressions (\ref{3_shish_lines}).}\label{ob22_shish}
\end{figure}

The figure \ref{ob22_shish} shows the phase diagram of such model. There are three regions, the green region where only the one cut solution exists, the blue region where the two cut solution exists and the red region, where also the asymmetric one cut solution exists. The green region extends all the way up for any value of $r>0$. There is however no stable solution under the line $h=-g$.

The three regions are separated by two parabolas, which are not difficult to compute. We take
$\reff=-1+4 h c_2$ and $\reff=-4\sqrt g$ or $\reff=-2\sqrt{15}\sqrt g$ for the upper or the lower boundary respectively. This together with the selfconsistency conditions (\ref{sec3_expressions1b},\ref{sec3_expressions1c}) yields the two boundary lines
\begin{subequations}\label{3_shish_lines}
\begin{align}
h(g)=&\frac{1}{4} \lr{\sqrt{g} - 4g}\ ,\\
h(g)=&\frac{3}{82} \lr{\sqrt{15} \sqrt{g} - 30 g}\ .
\end{align}
\end{subequations}
Both these expression precisely fit into the diagram \ref{ob22_shish} and the diagram we have obtained numerically nicely reproduces the analytical diagram from \cite{shishanin}. Note that the expression for the second line computed is different in this paper, probably due to some minor algebraic error.

\section{The second moment multitrace model - the asymmetric regime}\label{sec4}
In this section, we will analyze the asymmetric regime. This means, we will deal with the full action
\be S=\half F\lr{c_2-c_1^2}+\half r \trl{M^2}+g \trl{M^4}\ .\label{sec4_1}\ee
The saddle point equation for such model reads
\be -c_1 F'\lr{c_2-c_1^2}+r x_i + F'\lr{c_2-c_1^2} x_i+4 g x_i^4= \frac{2}{N}\sum_{i\neq j}\frac{1}{x_i-x_j}\ .\ee
This suggests, that the effective model to study is
\be S=a_{eff} \trl{M}+\half \reff\trl{M^2}+g \trl{M^4}\ .\label{sec4_4}\ee
We want to view this as an effective single trace model again. By rescaling the matrix $M$ in (\ref{sec4_1}) we can set the coefficient of the linear term to $1$ and obtain an effective model
\be S=\trl{M}+\half \reff\trl{M^2}+\geff \trl{M^4}\ ,\label{sec4_2}\ee
where
\be \reff=\frac{r+F'\lr{c_2-c_1^2}}{\lr{c_1 F'\lr{c_2-c_1^2}}^2}\ , \ \geff=\frac{g}{\lr{c_1 F'\lr{c_2-c_1^2}}^4}\ .\label{sec4_3}\ee
Both $r$ and $g$ are changed. So if we repeat the procedure that led to the figure \ref{fig1}, we expect a more complicated deformation of the phase diagram. We will need to keep in mind that the rescaling changes the relationship between the moments of the distributions in the effective single trace and the multitrace models.

We will now describe the phase structure of the model (\ref{sec4_2}) and then show, how it is deformed due to (\ref{sec4_3}) into a phase diagram of (\ref{sec4_1}).

\subsection{The effective asymmetric single trace model}

To keep the formulas legible, in this section we will omit the effective subscripts, understanding that the parameters are of the effective single trace model and not the multitrace model.

The phase structure of the effective model (\ref{sec4_2}) is more complicated and the full derivation of the phase diagram will be part of a different publication. Here, we will only describe the phase diagram, as we are mainly interested in the transformation of this diagram after inverting (\ref{sec4_3}). A more detailed analysis and discussion can be found in the review article \cite{mojeakty}.


\subsubsection*{Phases}

The model (\ref{sec4_2}) has two different phases. 

The one cut case, where the eigenvalue distribution is supported on a single interval ${(D-\sqrt{\delta},D+\sqrt{\delta})}$ and is given by
\begin{subequations}\label{sec4_1cut}
\be
\rho(x)=\frac{1}{2\pi}\lr{4 D^2 g + 2\delta g + r + 4 D g x + 4 g x^2}\sqrt{\delta-(D-x)^2}\ ,
\ee
with the conditions determining the boundary of the cut
\begin{align}
\half + 2 D^3 g + 3 D \delta g + \half D r=&0\ ,\label{sec4_1cut1}\\
-1 + 3 D^2 \delta g + \frac{3}{4}\delta^2 g + \quater \delta r=&0\ .\label{sec4_1cut2}
\end{align}
\end{subequations}

And the two cut case, where the support of the distribution is formed by two asymmetric intervals ${(D_1-\delta_1,D_1+\delta_1)}$ and ${(D_2-\delta_2,D_2+\delta_2)}$. The distribution is given by
\begin{subequations}\label{sec4_2cut}
\be
\rho(x)=\frac{2g}{\pi}\left|D_1+D_2+gx\right|\sqrt{\big(\delta_1^2-(D_1-x)^2\big)\big(\delta_2^2-(D_2-x)^2\big)}\ ,\label{sec41_doublerho}
\ee
with the conditions determining the boundary of the cuts
\begin{align}
2 D_1^2 g + 2 D_1 D_2 g + 2 D_2^2 g + \delta_1^2 g + \delta_2^2 g + \half r&=0\ ,\label{sec4_2cut1}\\
\half - 2 D_1^2 D_2 g - 2 D_1 D_2^2 g + 2 D_1 \delta_1^2 g + 
 2 D_2 \delta_2^2 g&=0\ ,\label{sec4_2cut2}\\
-1 + 2 D_1^2 \delta1^2 g - D_1 D_2 \delta_1^2 g - D_2^2 \delta_1^2 g + \quater\delta_1^4 g-&\nonumber\\ - D_1^2 \delta_2^2 g - D_1 D_2 \delta_2^2 g + 
 2 D_2^2 \delta_2^2 g - \half \delta_1^2 \delta_2^2 g + \quater\delta_2^4 &=0\ ,\label{sec4_2cut3}
\end{align}
and the condition
\be \int_{D_2+\delta_2}^{D_1-\delta_1}dx\,\rho(x)=0\ ,\label{sec4_2cut4}\ee
\end{subequations}
which chooses the two cut solution with the lowest free energy.\footnote{This formula looks innocent, but the explicit calculation with (\ref{sec41_doublerho}) yields an enormous formula involving elliptic integrals. For more details about this formula and its numerical treatment, see \cite{mojeakty}}

\subsubsection*{The phase diagram}

\begin{figure}%
\centering
\includegraphics[width=.7\textwidth]{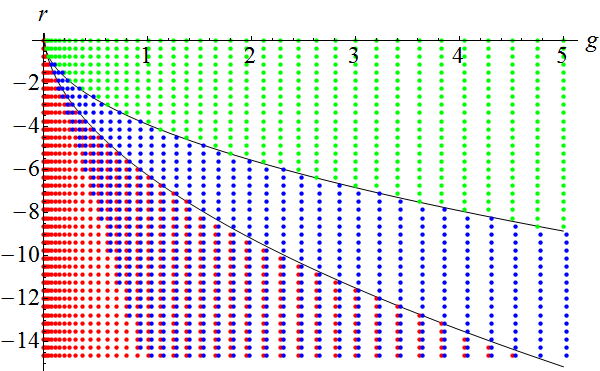}%
\caption{The phase diagram of the asymmetric single trace model (\ref{sec4_2}). The green region denotes the almost symmetric one cut phase, the red region the asymmetric one cut phase and the blue region the two cut phase. Note, that the dots for the two cut phase have been slightly shifted not to overlap with the red dots. The two lines denote the boundaries of existence (\ref{sec4_boundaries}).}%
\label{fig_asphase}%
\end{figure}

The conditions (\ref{sec4_1cut},\ref{sec4_2cut}) can not be solved analytically, so to obtain the phase diagram we need to proceed numerically \cite{mojeakty}. This means that we scan through the parameter space of the model and for particular values of $r$ and $g$ solve the equations numerically. If the solution exists, we make a little dot and move on. We obtain the phase diagram as shown in the figure \ref{fig_asphase}.

We have distinguished between two one cut solutions. In the almost symmetric solution the distribution extends over both of the minima of the potential, while the asymmetric one-cut solution lives completely in the left minimum of the potential.

We see that the region of existence of the two cut solution overlaps with the region of existence of the asymmetric one cut solution but not with the almost-symmetric one cut solution. By an explicit numerical computation, we find out that the two cut solution has lower free energy (\ref{3freenergy}) in the whole region where also the one cut solution exists and thus is the preferred solution.

Boundaries of the regions of existence of the one cut solutions can be computed perturbatively and are for the almost symmetric solution
\begin{subequations}\label{sec4_boundaries}
\be r(g)= - 4 \sqrt g+\frac{3}{32} + \frac{123}{32768 \sqrt g} + \frac{2343}{8388608 g}\ee
and for the boundary of existence of the asymmetric solution
\begin{align}
r(g)=&
- 2 \sqrt{15} \sqrt g
+ \frac{ 15^{1/4} g^{1/4}}{\sqrt 2}
+\frac{1}{12} 
+\frac{199}{5184 \sqrt 2 15^{1/4} g^{1/4}}+
\nonumber\\
&  
+ \frac{733}{ 62208 \sqrt{15} \sqrt g}
+\frac{49807}{ 5971968 \sqrt 2 15^{3/4} g^{3/4}} 
+\frac{2431}{ 11337408 g} +\nonumber\\
& 
+ \frac{244091717}{1393140695040 \sqrt 2 15^{1/4} g^{5/4}} 
\ ,\label{sec4_boundaries2}
\end{align}
\end{subequations}
which are the top and the bottom lines in the figure \ref{fig_asphase}. The boundary of existence of the two cut solution is difficult to compute, as it can not be obtained perturbatively. Fortunately, we will not need an explicit expression for this line in what follows.

In the next section, we will obtain the phase diagram of the asymmetric model (\ref{sec4_1}) as a deformation of the phase diagram in the figure \ref{fig_asphase}.

\subsection{The asymmetric multitrace model}

Obtaining the phase diagram of the asymmetric multitrace model (\ref{sec4_1}) is little more complicated than in the symmetric case. The reason is that due to the scaling of the matrix the moments of the distribution are not simply the moments of the effective distribution. Before, we did not need to distinguish between the moments of the two, but now we will have to make the distinction and make sure that we correctly translate between them. This makes the equations which determine the transformed phase diagram more involved.

We denote $x_0=-c_1 F'(c_2-c_1^2)$. To change the original action (\ref{sec4_4}) into (\ref{sec4_2}) we have to rescale the matrix $M\to M/x_0$. So the moments of the effective single trace model $c_{n,eff}$ are related to the moments of the distribution of the multitrace model by
\be c_1=\frac{c_{1,eff}}{x_0}\ ,\ c_2=\frac{c_{2,eff}}{x_0^2}\ .\ee
The value parameter $x_0$ is then determined by the self consistency condition
\be x_0=-\frac{c_{1,eff}}{x_0} F'\lr{\frac{c_{2,eff}-c_{1,eff}^2}{x_0^2}}\ ,\label{x0self}\ee
with the explicit formulas for the effective moments computable, but rather lengthy \cite{mojeakty}. For the fuzzy sphere this is a transcendental equation which is solved numerically.

We then obtain the transformation conditions
\be \reff=\frac{r+F'\lr{c_{2,eff}/x_0^2-c_{1,eff}^2/x_0}}{x_0^2}\ , \ \geff=\frac{g}{x_0^4}\ .\label{sec4_33}\ee
To obtain the phase diagram, we now proceed along the same lines. We scan through the parameter space $(\geff,\reff)$ by choosing some particular values, and then solve (\ref{x0self},\ref{sec4_33}) numerically to find the point $(g,r)$ in the phase diagram of the multitrace model (\ref{sec4_1}).

\begin{figure}%
\centering
\includegraphics[width=.7\textwidth]{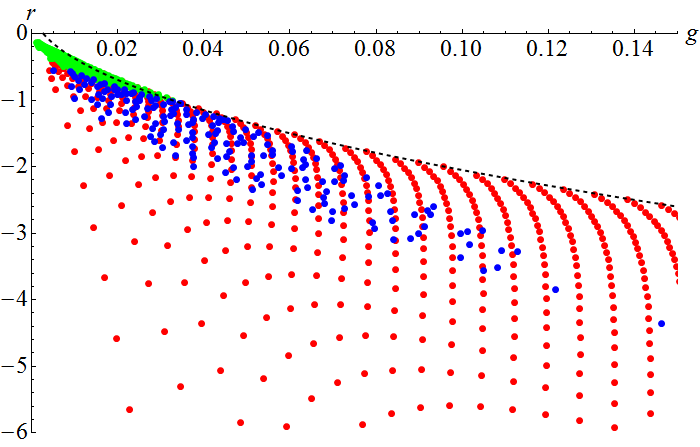}%
\caption{The phase diagram of the asymmetric multitrace model (\ref{sec4_1}). The green region denotes the almost symmetric one cut phase, the red region the asymmetric one cut phase and the blue region the two cut phase. The dashed line is the perturbative boundary of existence (\ref{sec4_boundpert}). Note a different scale on the $g$ axis. See the text for details.}%
\label{fig_asphasemulti}%
\end{figure}

The full phase diagram of the asymmetric regime of the theory is shown in the figure \ref{fig_asphasemulti}. The transformation of the diagram is now much more dramatic, the model does not have an asymmetric solution for large part of the parameter space and in the rest, only the asymmetric one cut solution is realized. The diagram also shows the perturbative expression for the existence of the asymmetric one cut solution computed in \cite{mojnovsi}
\begin{align}
r(g)=& 
- 2 \sqrt{15} \sqrt{g}
+\frac{2}{5}
- \frac{19}{18000 \sqrt{15} \sqrt{g}}
+ \frac{11}{150000 g}
+ \frac{44373739}{1458000000000 \sqrt{15} g^{3/2}}\no &
+ \frac{5033447}{6075000000000 g^2}
+ \frac{90528767950213}{248005800000000000000 \sqrt{15} g^{5/2}}  \no &
+ \frac{5907303225637}{516678750000000000000 g^3}
+ \frac{25212604606236508759}{4464104400000000000000000000 \sqrt{15} g^{7/2}} 
\ .\label{sec4_boundpert}
\end{align}
Clearly the line correctly describes the border of the transformed area of existence of the asymmetric one cut solution for larger values of $g$. For small values we run into similar problems as in the symmetric regime. It is also not difficult to check, that when we transform the boundary (\ref{sec4_boundaries2}) for the effective quantities into a boundary in terms of the original $r,g$ using (\ref{sec4_3}), we recover the formula (\ref{sec4_boundpert}).

\begin{figure}%
\centering
\includegraphics[width=0.85\textwidth]{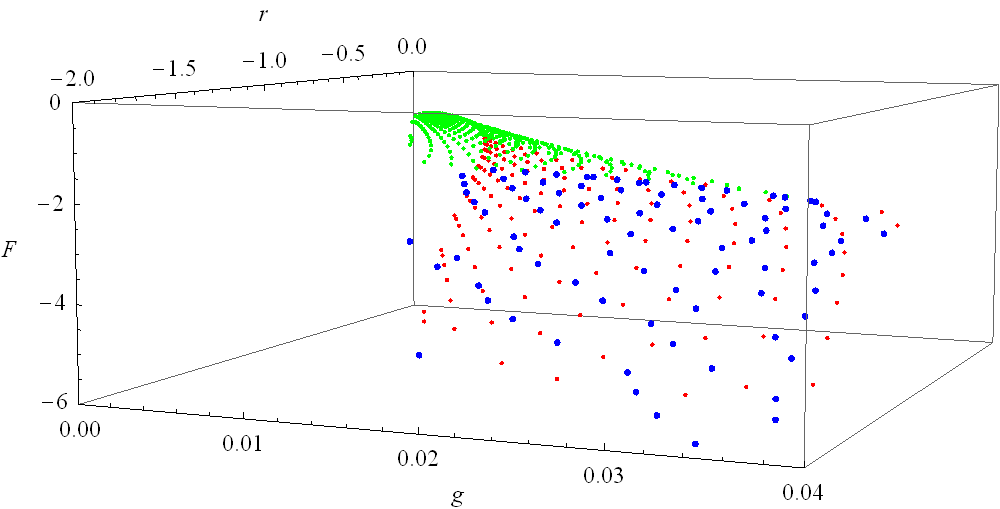}\\ %
\includegraphics[width=0.85\textwidth]{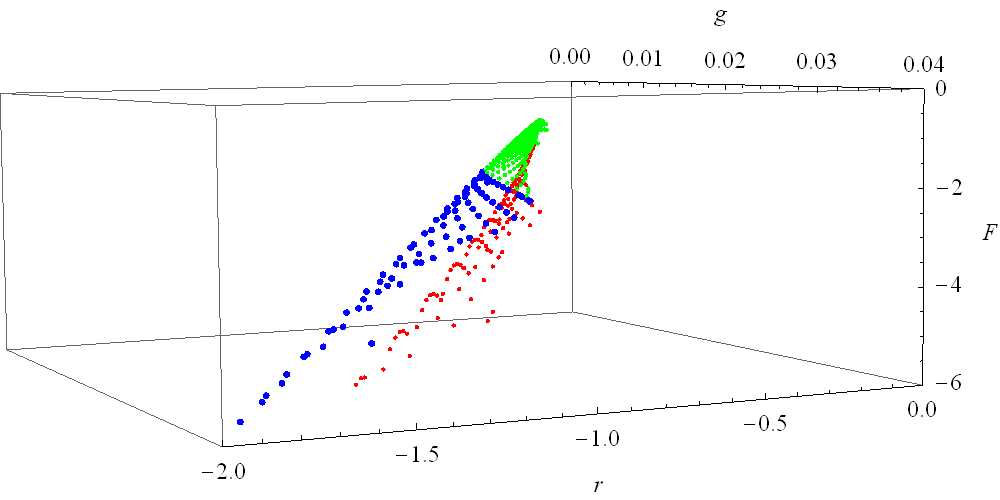}\\
\includegraphics[width=0.85\textwidth]{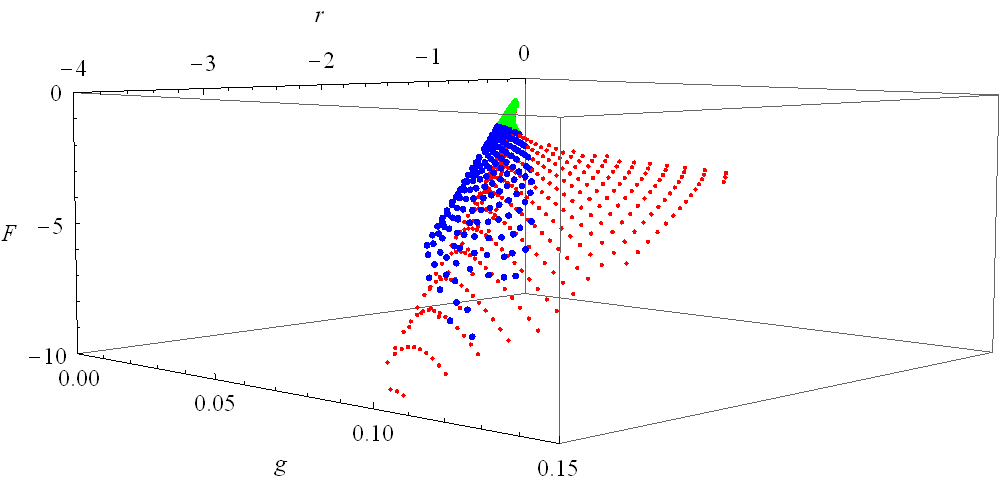}
\caption{The free energy diagram of the asymmetric multitrace model (\ref{sec4_1}). The green region denotes the almost symmetric one cut phase, the red region the asymmetric one cut phase and the blue region the two cut phase. Note the difference in scales between the first two and the last diagram. See the text for details.}%
\label{fig_asphasefree}%
\end{figure}

We see that in a region of parameter space more than one solution is available. The factor that determines which of the possible solution is realized is again the free energy, which we again compute by numerical integration of (\ref{3freenergy}). Opposing to the case of the effective model, the free energy of the asymmetric one cut solution is lower everywhere it exists and this is the preferred solution in the asymmetric regime, as is seen in the figure \ref{fig_asphasefree}.

This concludes the study of the asymmetric regime and we are ready to compare it with the symmetric regime.

\section{Interplay of the symmetric and asymmetric regimes in the second moment multitrace model}\label{sec5}
In this section, we shall put together the results of the previous two sections. If we assume that the fluctuations of the eigenvalues can change a symmetric distribution into an asymmetric one, for given values of $r$ and $g$, the system will be in the state with the lowest free energy.

\begin{figure}%
\centering
\includegraphics[width=1\textwidth]{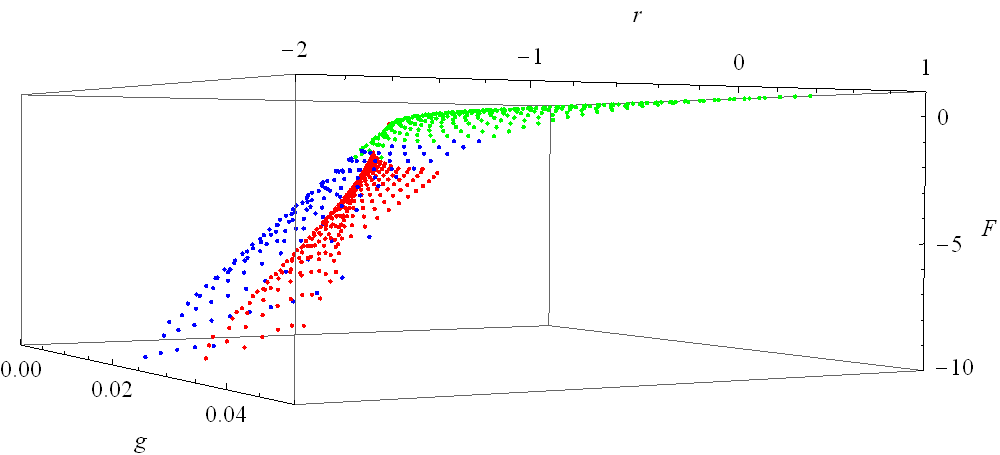}\\ %
\includegraphics[width=1\textwidth]{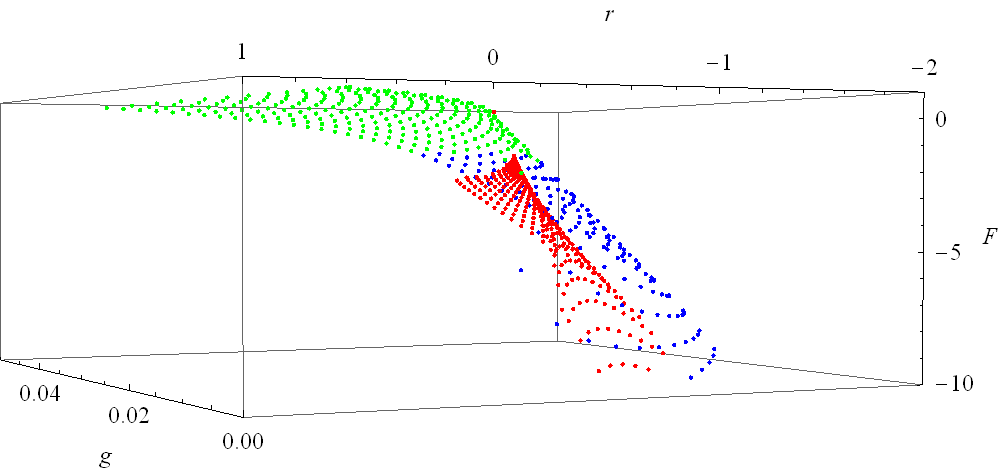}%
\caption{The complete free energy plot of the model (\ref{sec4_1}). The green, blue, red color code for one cut, two cut and the asymmetric one cut solutions respectively is kept. We see that the red region is always under the blue.}%
\label{fig_completefree}%
\end{figure}

\begin{figure}%
\centering
\includegraphics[width=.515\textwidth]{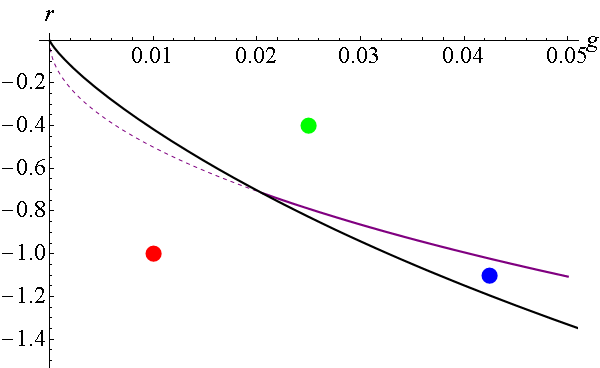}%
\includegraphics[width=.485\textwidth]{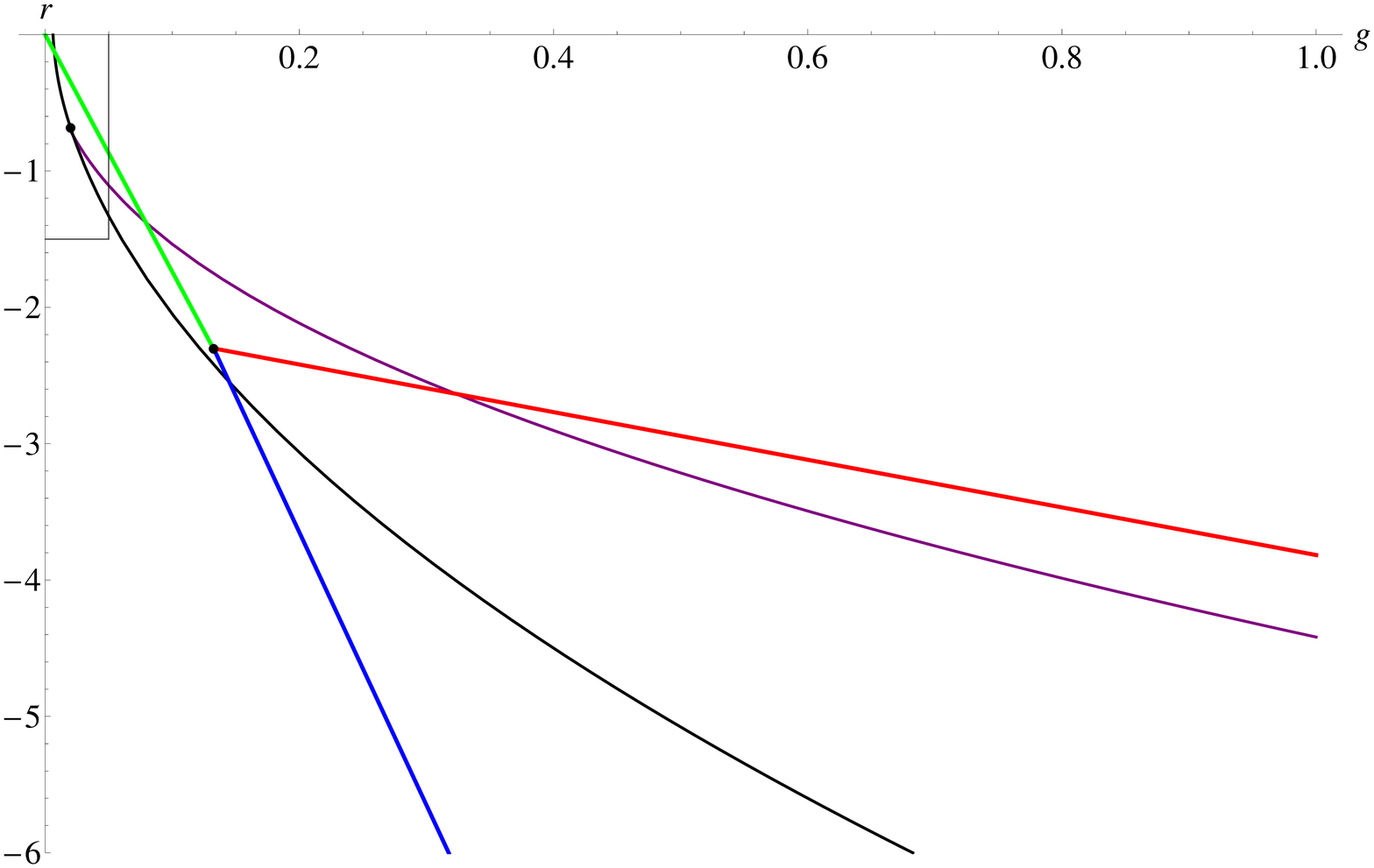}%
\caption{The first diagram shows the phase diagram of the model (\ref{sec4_1}). The green dot denotes the region of the symmetric one cut solution, or the disorder phase of the field theory. The red dot denotes the region of the asymmetric one cut solution, or the uniform order phase and the blue dot denotes the region of the two cut solution, or the nonuniform order phase. The second diagram shows a comparison of this phase diagram and the numerical phase diagram of the field theory. We show the extrapolation lines used to obtain the value of the triple point in the numerical work \cite{num09}. The rectangle in the top left corner shows the area covered by the phase diagram in the first figure.}%
\label{fig_completephase}%
\end{figure}

The complete free energy plot in the figure \ref{fig_completefree} reveals which one it is. We see that the asymmetric one cut solution is again the preferred solution everywhere it exists! All the other phases have larger free energy. We therefore obtain the phase diagram of the model (\ref{sec4_1}), describing the scalar field on the fuzzy sphere as in the figure \ref{fig_completephase}. The symmetric one cut solution corresponds to the disorder phase, the asymmetric one cut solution to the uniform order phase and the two cut solution to the nonuniform order phase.

The two boundaries in the figure are given by the exact expression (\ref{22trafo}) and the numerically computed boundary line of the red region from the figure \ref{fig_completefree}. These two meet at the triple point of the theory, for which we get the value
\be g_c=\ 0.02\ .\ee
Note, that the transition line from the critical point to the origin is the asymmetric one cut boundary and not the symmetric one cut boundary (\ref{22trafo}).

Obtaining this value was one of our main goals. It is not a perturbative result and does indeed take into account the whole structure of the kinetic term effective action $F(t)$. However we see, that the value is not too different from the perturbative one obtained in \cite{mojnovsi}.

The diagram we obtain shares the features and qualitatively agrees with the diagram obtained numerically, as can be seen from the second figure \ref{fig_completephase}. We obtain a lower value for the triple point than suggested from the Monte-Carlo simulations of the model \cite{num09,num14}. This has two possible explanations. One is clearly that the neglected terms in the remained $\mathcal R$ in (\ref{32sefffull}) deform the transition lines and shift the triple point.

However there is a possible imprecision also in the Monte-Carlo data. The transition lines are linear fits to the data points obtained quite fare from the triple point\footnote{For values of $g$ in the interval ${(0.4,0.6)}$.}. The second plot in \ref{fig_completephase} suggests that close to the origin the transition lines curve and the linear fit does not take this into account. So if the neglected terms do not change this drastically, the linear fits might overestimate the critical value $g_c$. Some preliminary numerical simulations closer to the origin of the phase diagram suggest that this is indeed the case, but at the moment it is not clear to what extent \cite{denjoepersonal}.

\section{Phase structure of the fourth moment multitrace models}\label{sec6}
Up to now, our primary aim was the nonperturbative second moment multitrace matrix model (\ref{2polymodel}), which involves only the powers of $\trl{M^2}$. As have been mentioned in the introduction, the first few terms of the perturbative expansion of $\mathcal R$ are known and we can try to estimate their effect.

These terms involve powers of $\trl{M^3}$ and $\trl{M^4}$ and will thus lead to a more complicated deformation of the phase diagram. There is a lot that can be said about similar multitrace models and we recommend the review \cite{mojeakty} for a thorough discussion. Here, we will proceed directly to the study of the matrix model corresponding to the fuzzy sphere scalar field\footnote{The qualitative results of this section change very little if one considers the first four terms of the $t$ expansion of $F(t)$, so we will consider this part in its entirety.}
\begin{align}
S(M)=&\half r\, \trl{M^2}+g\, \trl{M^4}+\frac{1}{2}F(c_2-c_1^2)-\frac{1}{432}\lr{c_3 - 3 c_1 c_2 + 2 c_1^3}^2-\nonumber\\&-\frac{1}{3456}\slr{(c_4 - 4 c_3 c_1 + 6 c_2 c_1^2 - 3 c_1^4)-2(c_2-c_1^2)^2}^2\ .\label{sec6_sphereAS}
\end{align}

\subsection{The symmetric regime}

\begin{figure} [tbp]
\centering 
\includegraphics[width=.7\textwidth]{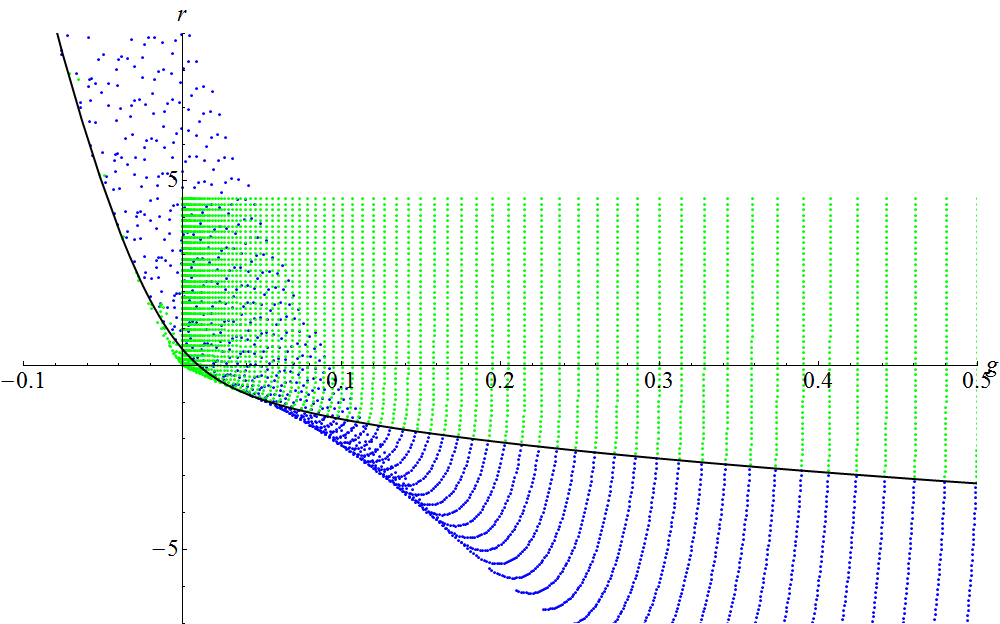}
\includegraphics[width=.7\textwidth]{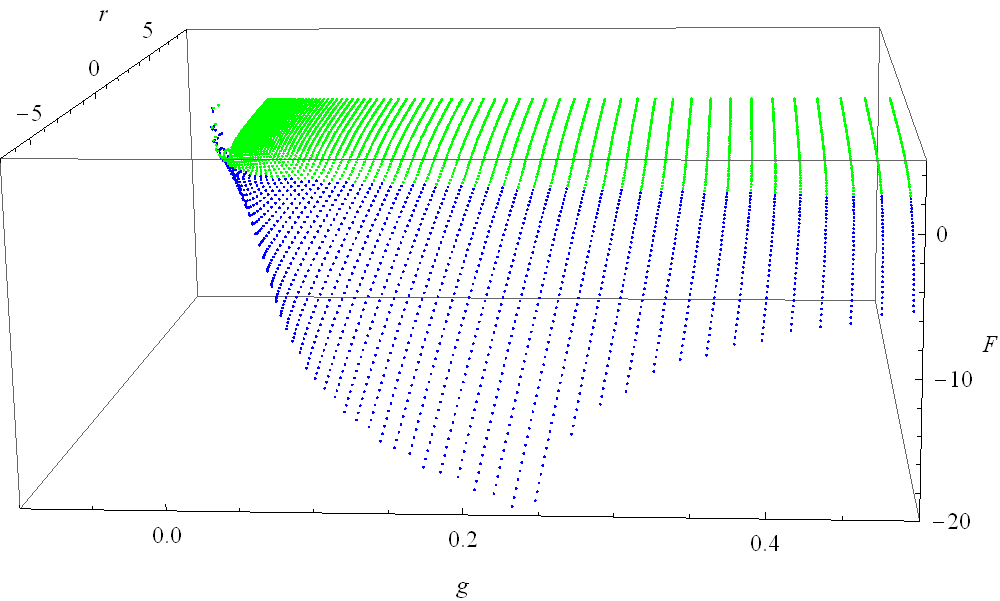}
\caption{The phase diagram and the free energy of the the matrix models (\ref{sec6_sphere}). The green region denotes the single cut solution, the blue region denotes the double cut solution. The deformation of the phase transition line (\ref{6_symtrafo}) is also shown.}\label{ob62_shere}
\end{figure}

After dropping the odd terms we obtain the symmetric regime action
\be S=\half r \trl{M^2}+g \trl{M^4}+\frac{1}{2}F(c_2)-\frac{1}{3456}\lr{c_4-2c_2^2}^2\ ,\label{sec6_sphere}\ee
The effective parameters are then given by
\be \reff=r+F'(c_2)+\frac{1}{216}c_2(c_4-2c_2^2) \ ,\ \geff=g-\frac{1}{1728}(c_4-2c_2^2)\ .\label{sec6_defomr}\ee
We repeat the same procedure as before, chose some values of $\reff,\geff$ and find the point in the $(g,r)$ diagram, to which it is deformed by (\ref{sec6_defomr}). There is a slight complication of having to consider the selfconsistency condition for the fourth moment and the change of $g$ also, but apart from that everything is rather straightforward. The resulting phase diagram and the free energy diagram of the model are shown in the figure \ref{ob62_shere}.

It is again not difficult to compute the transition line of this model, since at the phase transition the expression simplify. The resulting expression is
\be\label{6_symtrafo}
r(g)=-\frac{1}{1 - e^{\frac{4 \sqrt 3}{\sqrt{24 g + \sqrt{1 + 576 g^2}}}}} 
 +\frac{2}{\sqrt 3 \lr{24 g + \sqrt{1 + 576 g^2}}^{3/2}}
 - \frac{5 \sqrt{24 g + \sqrt{1 + 576 g^2}}}{4 \sqrt 3}
\ee
and is also shown in the figure \ref{ob62_shere}. We stress that this formula is an exact transition line for the model (\ref{sec6_sphere}), without any perturbative considerations. However if the model is viewed as an approximation of the fuzzy field theory, the action is a perturbative expansion and the given formula is just an approximation of the true transition line.

And quite clearly it is not a good approximation. The phase diagram has a lot of features which we do not expect in the phase diagram of the field theory. It has no stable solution in a large part of the parameter space, in some parts of the parameter space it allows for two different symmetric solutions, it has a stable configuration also for $g<0$ and the phase transition line does not go through the origin of the parameter space.

\subsection{The asymmetric regime}

Unfortunately there is little we can do with our approach in the case of the full model (\ref{sec6_sphereAS}), since the model is too complicated to analyze \cite{mojeakty}. The root of the problem is that to turn the model into an effective single trace model of the form (\ref{sec4_2}), on top of the scaling one has also to shift the matrix, which brings an untraceable dependence of $r,g$ on the effective moments.

To include the effect of the new terms in some way, we repeat the procedure of \cite{mojnovsi} to compute the perturbative expansion of the transition line and obtain
\be r(g)= - 2 \sqrt{15} \sqrt{g}+\frac{2}{5} - \frac{19}{ 18000 \sqrt{15} \sqrt{g}} 
+\frac{29}{1125000 g}
- \frac{7886183}{4374000000000 \sqrt{15} g^{3/2}}
\ .\ee
We stress that this formula is a perturbative one even if the model (\ref{sec6_sphereAS}) is viewed as a complete action. This transition line does not intersect with the transition line (\ref{6_symtrafo}), which is again something which should not happen.

We thus find out that the perturbative model (\ref{sec6_sphereAS}) is not a good approximation and does not describe the scalar field on the fuzzy sphere well. To include the effect of higher moments one has to find a nonperturbative way to include $t_3,t_4$, or perhaps a complete treatment of the integral (\ref{31anfintegral}). Taking only the first few terms is not going to work, the fuzzy field theory is very sensitive to the terms of the perturbative expansion and to be able to make any reasonable statements about the properties of the phase diagram of the field theory, we need to include terms of all orders.


\section{Conclusions}
We have studied two different multitrace matrix models which approximately describe the scalar field theory on the fuzzy sphere.

The first model was nonperturbative and included terms of all orders in the first and second moments of the eigenvalue distribution. We have shown that this model leads to a phase diagram that does have the same features as the phase diagram obtained by Monte-Carlo simulations. The triple point of the phase diagram is however lower from the numerical value by a factor of $7$. It has been argued that there is a room for improvement in both analytical and numerical approaches.

The second model we studied was a perturbative one, including some terms involving the fourth moment of the distribution. However it has been shown, that this model leads to a phase diagram with features very different from the expected field theory diagram. We have concluded that the phase diagram is very sensitive to the perturbative nature of the model and any reliable treatment of the phase diagram has to be nonperturbative. Let us stress one final time, that this was not due to a perturbative treatment of the model, but due to the fact that the model itself was a perturbative approximation. A similar result has been obtained recently by a direct numerical simulation of the multitrace matrix model \cite{ydrisupernew}. It would be interesting to see, how our results connect to this work.

It would also be interesting to see, if there is a more complete, or perhaps the complete treatment of the angular integral (\ref{31anfintegral}). At the moment is not clear how to approach this problem, as the perturbative treatment seems to be way too complicated \cite{saman2} and there is no hint how to generalize the nonperturbative treatment of \cite{poly} beyond the second moment.

On the other hand, there is a lot of numerical data for different fuzzy and noncommutative spaces, which one could try to reconstruct with the presented tools. The fuzzy disc \cite{fuzzydiscnum}, three dimensional case of ${\mathbb R\times S^2_F}$ \cite{rsfnum}, noncommutatie plane \cite{num14panero2} to name some. It would be interesting to see how well this method does in all these cases, or perhaps to give some predictions for possible numerical study of theories on different spaces, such as $\mathbb CP^2,\mathbb CP^3$ or fuzzy $S^4$ \cite{s4}.

Finally, the modifications of field theories with no UV/IR mixing remain the most interesting ones to study \cite{uvir,gw,nouvir2}. As we have mentioned, the nonuniform phase in the commutative limit of the phase diagram is a consequence of the UV/IR mixing. In the modified phase diagram this phase should thus be absent. This is something that remains to be shown and perhaps one could learn something new from the way the extra phase is removed.

\acknowledgments
I would like to thank Denjoe O'Connor for useful discussions.
This work was supported by the \emph{Alumni FMFI} foundation as a part of the \emph{N\'{a}vrat teoretikov} project.
Part of this work was done while I have been visiting the Erwin Schr\"odinger International Institute for Mathematical Physics in Vienna for "The interrelation between mathematical physics, number theory and noncommutative geometry" workshop.
This article is based upon work from the COST Action MP1405 QSPACE, supported by COST (European Cooperation in Science and Technology), which also supported my stay at the "EISA Workshop on Noncommutative Field Theory and Gravity", where this work was presented.




\end{document}